# Automatic Volumetric Segmentation of Additive Manufacturing Defects with 3D U-Net

**Vivian Wen Hui Wong,[1] Max Ferguson,[1] Kincho H. Law,[1] Yung-Tsun Tina Lee,[2] Paul Witherell[2]**

[1]Engineering Informatics Group, Civil and Environmental Engineering, Stanford University, Stanford, CA 94305
[2]Systems Integration Division, National Institute of Standards and Technology (NIST), Gaithersburg, MD 20899
[1]{vwwong3, maxferg, law}@stanford.edu; [2]{yung-tsun.lee, paul.witherell}@nist.gov

**Abstract**

Segmentation of additive manufacturing (AM) defects in X-ray Computed Tomography (XCT) images is challenging, due to the poor contrast, small sizes and variation in appearance of defects. Automatic segmentation can, however, provide quality control for additive manufacturing. Over recent years, three-dimensional convolutional neural networks (3D CNNs) have performed well in the volumetric segmentation of medical images. In this work, we leverage techniques from the medical imaging domain and propose training a 3D U-Net model to automatically segment defects in XCT images of AM samples. This work not only contributes to the use of machine learning for AM defect detection but also demonstrates for the first time 3D volumetric segmentation in AM. We train and test with three variants of the 3D U-Net on an AM dataset, achieving a mean intersection of union (IOU) value of 88.4%.

## 1 Introduction

Additive manufacturing (AM), also known as three-dimensional (3D) printing, is experiencing rapid adoption in the manufacturing domain. AM introduces several advantages such as the fabrication of models with complex geometries, quick prototyping, customization of materials and flexibility in design (Ngo et al., 2018). During an AM fabrication process, consecutive layers are printed (Gross et al., 2014). Internal defects can be created due to reasons such as print error, residual stress, or cyber-attack (Holzmond and Li, 2017). The presence of defects leads to flaws, such as insufficient material properties, in the printed object (Reese, Bheda and Mondesir, 2016). An automatic defect segmentation system to identify interior defects can therefore aid AM quality control.

Some current non-destructive defect segmentation techniques include infrared radiation monitoring and X-ray computed tomography (XCT). The former monitors radiation given off by melt pools during the printing process. The monitoring process requires a complex laser setup and only works for the powder bed fusion (PBF) process (Holzmond and Li, 2017). XCT can be used to visualize internal structures, including porosity, in three dimensions (Buffiere et al., 2001; Masad et al., 2002). While XCT can be used to obtain images and segmentation labels, it requires manual thresholding, which can be tedious given many samples. Other approaches including analyzing an array of sensor signals to monitor the process (Rao et al., 2015), but such approaches require the installment of multiple sensors and analysis of multiple signal types. Automatically identifying small defects inside an object therefore remains a challenging task in manufacturing.

Defect segmentation can be treated as an image segmentation problem in computer vision. In defect segmentation, each 2D pixel or 3D voxel is classified as either defect or background. Image segmentation is a difficult task, and many methods have been proposed to solve the problem, especially in the medical imaging domain, which requires localization of objects (Ronneberger, Fischer and Brox, 2015). State of the art 2D and 3D segmentation methods that perform best in various segmentation tasks are based on deep learning and use convolutional neural networks (CNN) (Guo et al., 2018). The benefits and the drawbacks of some of the current segmentation methods are further discussed in Section 2.

Despite their drawbacks, 3D CNNs have demonstrated potential in 3D medical image segmentation tasks (Cicek et al., 2016; Milletari, Navab and Ahmadi, 2016; Lee et al., 2017; Yu et al., 2017; Zhou et al., 2018; Ghavami et al., 2019). AM images share similar characteristics with medical images such as their volumetric nature and similar level of contrast. Therefore, it is thought that certain deep learning approaches from the biomedical domain might be of similar benefit in the AM domain. However, processing XCT data from AM parts poses certain difficulties that are less prevalent in the biomedical domain, such as small and highly irregular defect geometries. In addition, the sparsity of defects in AM varies dramatically, as exemplified by the dataset used in this paper which contains samples that range from 0.37% to 19.38% porosity. Furthermore, AM datasets are both difficult and costly to produce and hence there are very few publicly available datasets large enough to employ machine-learning approaches. Despite these challenges, developing fast and reliable quality control procedures is essential to the mainstream adoption of additive manufacturing processes.

The need for automatic manufacturing defect detection and the promising performance of CNN in the volumetric segmentation of medical images thus motivates our work. We propose to train a 3D U-Net model with existing defect labels and use the trained model to automate the segmentation process on XCT or 3D images of unknown additive manufacturing samples. To demonstrate the effectiveness of our method, an experiment is conducted using an AM defect dataset constructed with XCT images. We train and evaluate 3 variants of the 3D U-Net model and obtain a mean intersection of union (IOU) value of 0.863 to 0.884. Although no previous work has been done in AM defect segmentation to compare this accuracy with, our mean IOU is comparable with the accuracy of 3D U-Net segmentation of kidney embryos, which achieves a mean IOU of 0.704 (Cicek et al., 2016).

The rest of the paper is organized as follows. Section 2 provides an overview of related works. Section 3 briefly introduces 3D CNNs and presents the network architecture of our 3D U-Net models. Section 4 discusses our implementation and experimental results. Finally, Section 5 briefly concludes our work.

## 2   Related Works

This section reviews current advancements in 3D image segmentation using deep learning approaches, as well as related works in automatic AM defect identification.

Segmentation of 3D objects are often framed as a 2D segmentation task with post processing (Zhou et al., 2016; Milletari, Navab and Ahmadi, 2016). The most commonly used CNN architectures used for 2D segmentation problem are region-based and fully-convolutional-network-based (FCN-based) (Guo et al., 2018).

Region-based segmentation methods first extract regions and describe them, then classify the region (Caesar, Uijlings and Ferrari, 2016). Region-based CNN (R-CNN) is a representative method using this approach (Guo et al., 2018). The Mask R-CNN architecture is an example of R-CNN that performs object detection and segmentation simultaneously (He et al., 2017). Mask R-CNN has proven to be effective in segmenting everyday objects (He et al., 2017), medical images (Johnson, 2018) and metal casting defects in manufacturing (Ferguson et al., 2018).

On the other hand, FCN-based methods perform segmentation by directly learning a mapping from input to output pixels, without proposing regions (Long, Shelhamer and Darrell, 2015). U-Net is a CNN model that extends the FCN architecture, achieving excellent performance in the segmentation of ventral nerve chord (Ronneberger, Fischer and Brox, 2015). U-Net is less computationally expensive than Mask R-CNN, since it does not require the generation of region proposals.

Despite the success of the above-mentioned methods, AM specimens are 3D, and therefore predicting one 2D slice at a time loses information on the correlation between slices (Yu et al., 2019). A similar problem exists in the medical imaging domain, where most images are 3D volumes. To that end, 3D CNN models have been developed to make predictions on volumetric medical images (Cicek et al., 2016; Milletari, Navab and Ahmadi, 2016).

While 3D CNN models can leverage information between slices, they lack pre-trained models, leading to less stable training (Yu et al., 2019). Patch-wise predictions are also more time-consuming to generate compared to 2D predictions. V-Net (Milletari, Navab and Ahmadi, 2016) and 3D U-Net (Cicek et al., 2016) are examples of end-to-end architectures for 3D segmentation. In this work, we adopt 3D U-Net and two of its variants in our experiments with AM defect data.

A few related works have been done to automatically detect AM defects using CNN. Scime and Beuth (2018) and Zhang, Liu and Shin (2019) perform 2D detection and classification on defects using slices of camera images in their work. Shevchik et al. (2018) use CNN to analyze acoustic emissions during AM processes.

As discussed in this section, the use of 3D CNN to segment AM defects represents a new and novel approach, and will be the subject of discussion in the rest of this paper.

## 3   3D Convolutional Neural Networks

In this section, we present some background knowledge on 3D CNNs. As discussed in the previous section, 2D CNNs and 3D CNNs each have their own drawbacks and advantages. 3D CNNs extend upon 2D CNNs by using the same technique of convoluting a kernel spatially through the input of a convolutional layer with two important distinctions:

1) The kernel of a 2D convolutional layer is two dimensional (2D) with width and height ($W \times H$), and the kernel of a 3D convolutional layer is three dimensional (3D) with width, height and depth ($W \times H \times D$).
2) A 2D kernel moves in 2 directions, along the axis corresponding to W and H dimensions. A 3D kernel moves in 3 directions, with an additional axis along the D dimension.

**3D U-Net**

The 3D U-Net is an extension upon the standard (2D) U-Net architecture proposed by Ronneberger, Fischer and Brox,

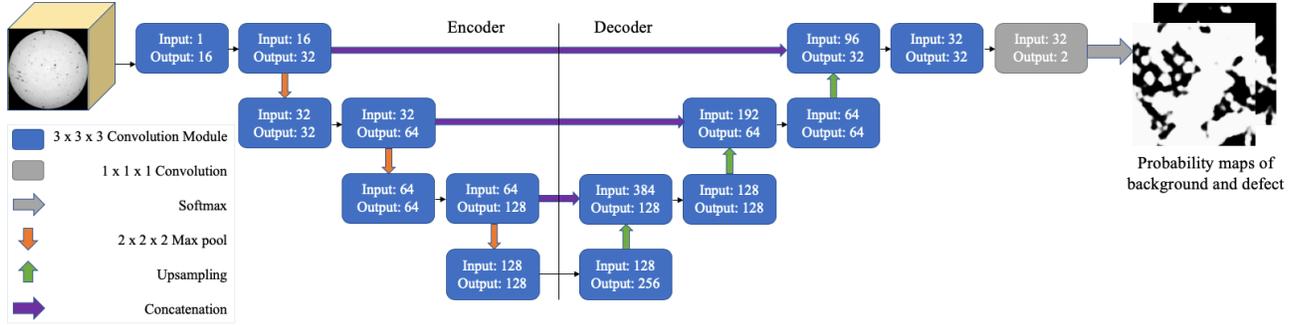

Figure 1: 3D U-Net architecture implemented in this study. The numbers on the convolutional blocks indicate the number of input and output channels (Cicek et al., 2016).

Table 1: Details of the AM defect data

| Specimen | Distance between 2D slices [pixel] | Total number of voxels | Porosity (%) | Dataset |
|---|---|---|---|---|
| Sample 1 | 0.00245 | $6.57 \times 10^8$ | 1.01 | Validation |
| Sample 2 | 0.00277 | $6.61 \times 10^8$ | 19.28 | Training |
| Sample 3 | 0.00243 | $6.55 \times 10^8$ | 0.37 | Training |
| Sample 4 | 0.00252 | $5.50 \times 10^8$ | 11.01 | Training |

(2015). The 3D U-Net can be implemented using a modular architecture, as shown in Figure 1. The architecture consists of two types of modules, encoder modules shown on the left and decoder modules on the right. The encoder modules perform max pool operations and the number of feature maps produced increases as the number of layers increases. On the other hand, the decoder modules perform upsampling and the number of feature maps produced decreases as the number of layers increases. This overall encoder-decoder design is preserved for all 3D U-Net models, but the design of each module can be altered. Each 3 x 3 x 3 convolution module consists of a convolutional layer as well as group normalization (GN) or batch normalization (BN) layer, and a rectified linear unit (ReLU) layer (Cicek et al., 2016).

Both GN and BN are techniques to improve the speed and stability of training neural networks (Ioffe and Szegedy, 2015). Wu and He (2018) propose GN, which is more robust than BN, and does not suffer the limitation of BN that smaller batch size leads to larger errors. The 3D U-Net in the work by Cicek et al. (2016) uses BN and reports outstanding results. To conduct the experiments on the AM defects, the 3D U-Net implementation by Cicek et al. (2016) is employed. However, we modify the convolutional modules to replace BN with GN, as well as rearranging the layer structures, allowing us to achieve a higher accuracy on the AM defects, as discussed in the next section.

Residual Symmetric 3D U-Net is an improved variant of 3D U-Net, proposed by Lee et al. (2017). Several changes are made compared to the 3D U-Net architecture, including the addition of a layer by redesigning convolutional modules to include residual skip connections and symmetricity, as well as modifying the downsampling and upsampling techniques. Lee et al. (2017) report that their model exceeds human accuracy in an experiment segmenting neurites in electron microscopic (EM) brain images.

As discussed in the next section, we modify the convolution module of the 3D U-Net architecture to analyze their impact on the model's performance on the AM defect segmentation task. We also compare Residual Symmetric 3D U-Net against the two designs.

## 4 Implementation Details and Experimental Results

This section describes the implementation and results of the experiments conducted using a dataset of AM defect images.

### 4.1 Data

The dataset consists of four cylindrical AM specimens, shown in Table 1. The datasets were introduced and analyzed by Kim et al. (2017), and the data are publicly available (Kim et al., 2019). The artificial defects were produced by changing AM processing parameters. The sample contains AM defects due to unoptimized AM processing parameters. Each specimen consists of 8-bit grayscale images of 2D slices. These images are 16-bit raw images obtained using XCT reconstruction processed by adding a median 3D filter and a non-local means filter (Buades, Coll and Morel, 2011; Sun, Brown and Leach, 2012). To obtain the segmentation mask, the 8-bit images are processed with Bernsen local thresholding (Bernsen, 1986). The local contrast threshold parameters of the thresholding process are set by relating average noise value to local contrast threshold as explained by Kim et al. (2017). Examples of some images and corresponding masks are shown in Figure 2.

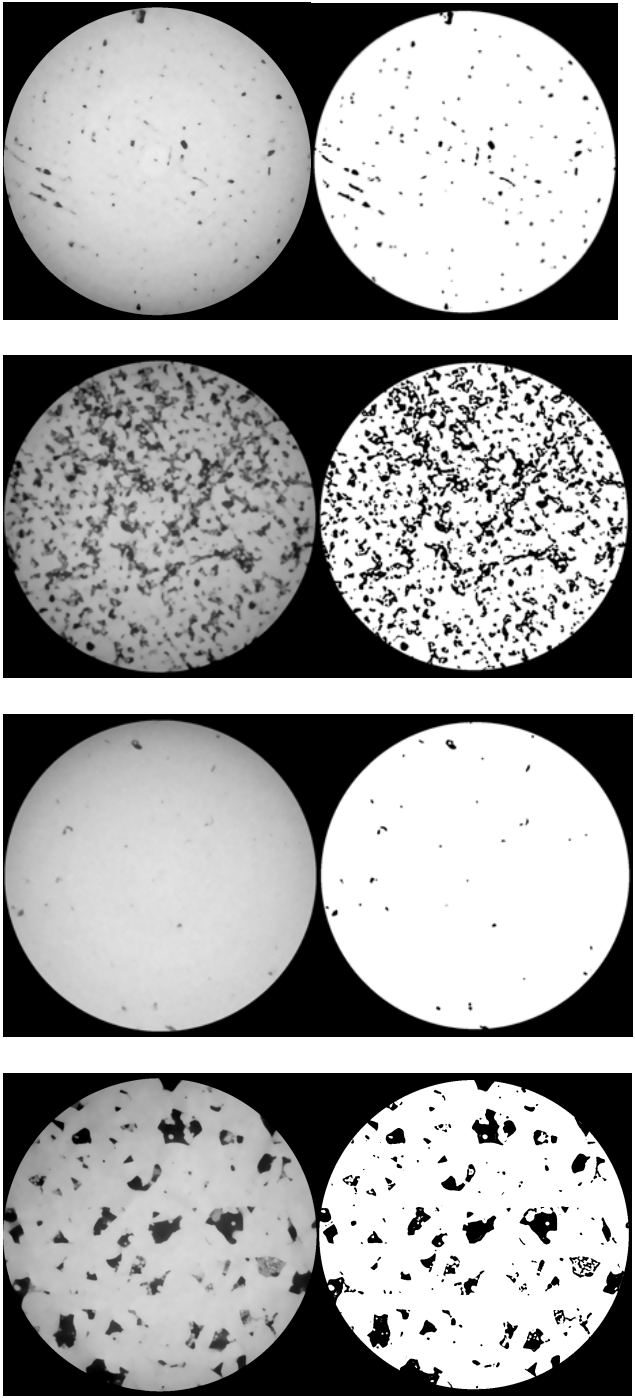

Figure 2: Examples of images from the AM defect dataset. Processed XCT images are on the left and segmentation masks are on the right.

In the experimental study, Sample 1 is used for validation. Samples 2, 3 and 4 are used for training. This choice is due to the fact that Sample 1, among all samples, does not have an extreme value of porosity.

### 4.2 Training and Inference

To analyze the effect of 3D U-Net convolutional modules on performance and convergence, three 3D U-Net models are trained and evaluated in this study. The three models vary in the layers of the basic convolutional modules. The first model, hereby referred to as 3D U-Net with Conv+BN+ReLU, follows the implementation by Cicek et al. (2016). The model uses a 3D convolutional layer, a BN layer and a ReLU nonlinearity layer in its basic convolutional module. The second model, 3D U-Net with Conv+ReLU+GN, is a variant that uses a 3D convolutional layer, a ReLU nonlinearity layer, followed by a GN layer. The third model, Residual Symmetric 3D U-Net, follows the implementation by Lee et al. (2017).

Network inputs are 3D images of dimensions Depth × Width × Height, constructed by stacking the 2D slices, as shown in Figure 3. The inputs are normalized, randomly flipped and rotated prior to training. Network outputs and targets are compared using the softmax function with cross-entropy loss. The models are trained end-to-end and without pretraining for 2000 iterations using a NVIDIA Tesla T4 GPU, which fits patches of size 128 × 128 × 128.

The models are fine-tuned on the AM defect dataset. Each model is trained with an initial learning rate of 0.0002 that decays at a rate of 0.5 at the 600th, 1000th, and 1400th iteration. The networks are trained via the Adam optimizer (Kingma and Ba, 2014). A weight decay factor of 0.0001 is used. We set the batch size and the group size of one in BN and GN, respectively. Stride sizes are 32 × 32 × 32 to overlap the patches and to capture the fine details in the images. All modifications and fine-tuning to the models are conducted using a publicly available implementation of the 3D U-Net architecture (Wolny, 2019).

The models are evaluated with the same GPU. The prediction accuracy of each model is evaluated using the mean intersection over union (IOU) metric.

### 4.3 Results and Discussion

The performance of the three different 3D U-Net models are compared in Table 2. Each model's validation accuracy and the amount of time taken to achieve that are reported in the table. The Residual Symmetric 3D U-Net model, with a mean IOU of 0.884, exceeds the other models slightly in performance, but requires the longest training time. With the 3D U-Net model, GN improves the result with respect to BN, but is slower in the training iterations. In general, we observe a trade-off between training time and accuracy. In practice, an appropriate model should be selected taking into consideration the time and accuracy trade-off.

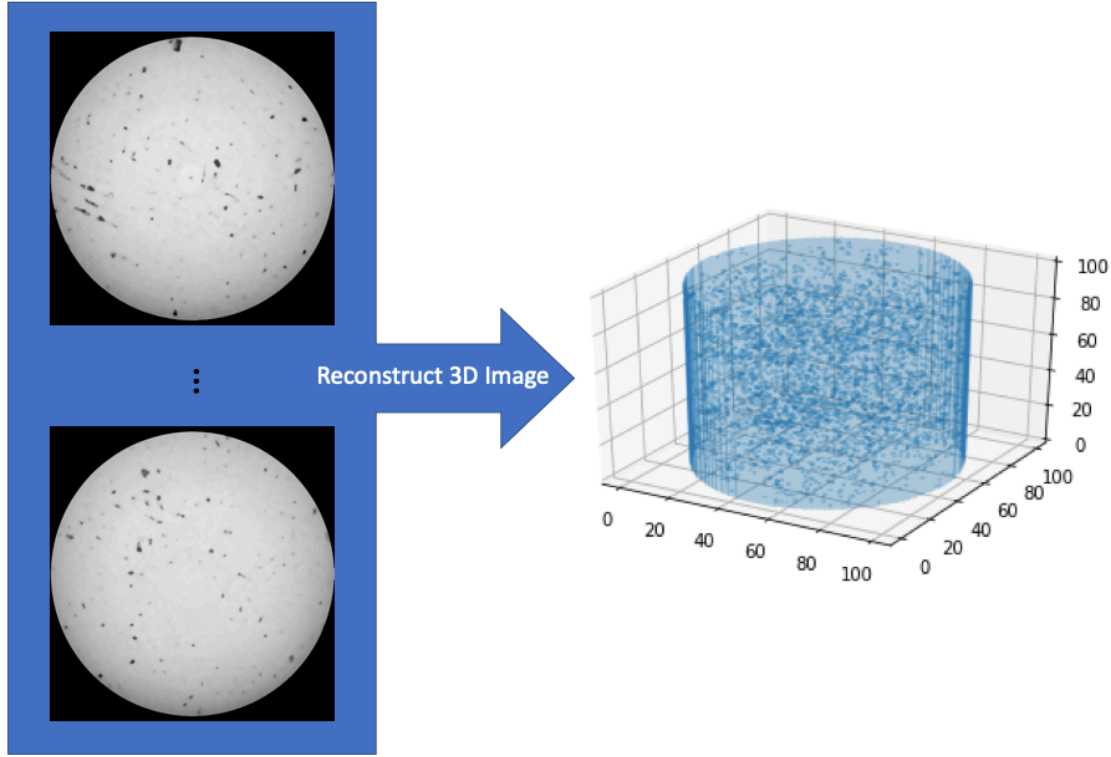

Figure 3: Reconstruction of a 3D image. Note that the voxels on the right are downsized for display purpose.

Table 2: Mean IOU and average training time

| Model | Training Time on GPU [hours] | Validation Mean IOU |
|---|---|---|
| 3D U-Net with Conv+BN+ReLU | **6.58** | 0.863 |
| 3D U-Net with Conv+ReLU+GN | 14.00 | 0.881 |
| Residual Symmetric 3D U-Net | 19.97 | **0.884** |

Figure 4 shows an example of a slice segmented using the Residual Symmetric 3D U-Net model. It can be seen that the segmentation probability map compares well with the target (labelled) sample. One observation is that the voxels corresponding to sharp geometries of the defects are often misclassified. Due to the size of the defects and poor contrasts at the edges, these sharp geometries appear to be small and light in color, and their true label is often ambiguous, hence posing difficulties for segmentation.

## 5 Summary and Discussion

This paper has presented a method for automatic volumetric segmentation of AM specimens using 3D U-Net, a CNN model previously developed for medical image segmentation. Three variants of the model are compared using an AM defect dataset, and the highest mean IOU achieved is 0.884, which is a good accuracy considering the various challenges in segmenting small defects. The proposed method is able to automatically segment defects in AM samples with a reliable amount of accuracy, and can be of assistance to quality control for the additive manufacturing process.

Future work could focus on tuning the network to handle the misclassification of areas with very few voxels. One improvement might involve using focal loss, which puts less penalty on well classified samples and focuses on misclassified samples (Lin et al., 2017). Dilation could be used to pre-process the training images to add voxels to the boundaries of the defects (Jackway and Deriche, 1996). The proposed models could also be trained to segment defects in AM specimens with different geometries, materials and additive manufacturing approaches. Furthermore, the effect of transfer learning on the performance and training speed of the CNN models could be explored. It would also be interesting to develop CNN models to focus on studying

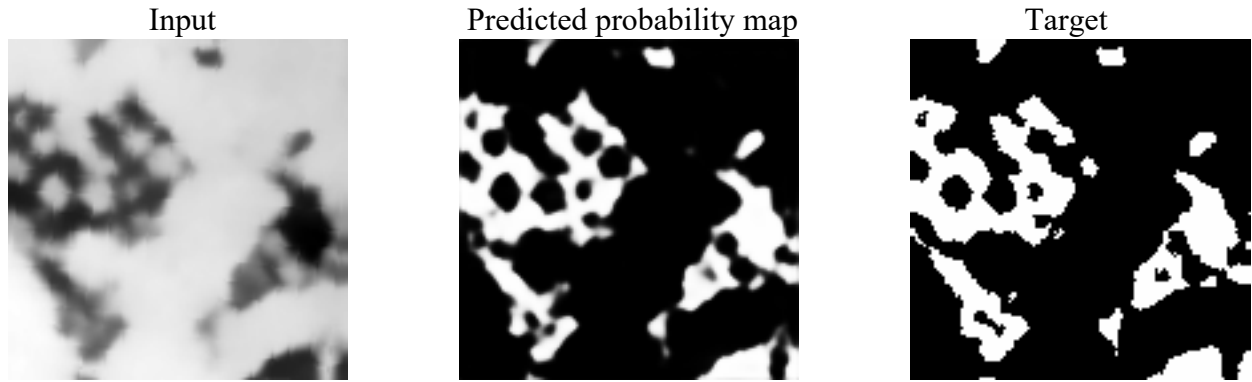

Figure 4: An example slice of a defect and its segmentation probability map outputted by Residual Symmetric 3D U-Net.

other important characteristics of AM defects, such as defect pattern classification.

The ability to segment defects with high accuracy could be beneficial in a number of situations in practice. A trained CNN model could be used to evaluate fabricated AM specimen for quality assurance. Furthermore, images of a specimen could be captured in the middle of fabrication to identify warning signs early on during the printing process, thereby saving materials from a failed printing process. However, to deploy this method in practice would require more training data and ensure that the model generalizes to other types of AM defects.

## Acknowledgement and Disclaimer

This research is partially supported by the Smart Manufacturing Systems Design and Analysis Program at the National Institute of Standards and Technology (NIST), US Department of Commerce, Grant Number 70NANB19H097 awarded to Stanford University. The authors would like to thank Dr. Felix Kim of NIST for providing the 3-D printing data sets used for the experimental study.

Certain commercial systems are identified in this article. Such identification does not imply recommendation or endorsement by NIST; nor does it imply that the products identified are necessarily the best available for the purpose. Further, any opinions, findings, conclusions, or recommendations expressed in this material are those of the authors and do not necessarily reflect the views of NIST or any other supporting U.S. government or corporate organizations.